\documentstyle[twocolumn,aps,prl,epsf]{revtex}

\begin{document}
\draft
%\title{ Cascades of Transitions
%and Classical-Quantum Correspondence}
\title{ $\hbar \to 0$ in Kicked Harper Model: Reassurances and Surprises}
\author{Indubala I. Satija$^a$ and Toma\v z Prosen$^b$  }
\address{
(a)Department of Physics, George Mason University, Fairfax, VA 22030\\
(b) Physics Department, Faculty of Mathematics and Physics,
Univ.of Ljubljana, Jadranska 19, SI-1111 Ljubljana, Slovenia}
\date{\today}
\maketitle
\begin{abstract}
We investigate classical-quantum correspondence for kicked Harper model 
for extremely small values of the Planck constant $\hbar$.
In the asymmetric case a pure quantum state shows clear signature of
classical diffusive as well as super diffusive transitions asymptotically 
independent of $\hbar$.
However, for the symmetric case, the $\hbar$ independent behavior 
occurs only for renormalized parameter $\bar{K}=K/(2\hbar)$ with intriguing 
features such as a sharp transition from integrable to non-integrable 
transport at $\bar{K}=\pi/2$, a series of transitions at multiples of 
$\pi$ and periodicity of the transmission probability. 
These results add new puzzles to the frontiers of quantum chaos.
\end{abstract}

\narrowtext

Localized transport of quantum system in the regime where the corresponding 
classical system exhibits deterministic diffusive behavior is one of the most 
surprising aspects of non-integrable Hamiltonian systems \cite{CC}.
However, the correspondence principle requires some signatures of various 
classical transitions such as the breakup of 
KAM tori leading to diffusive transport and the emergence of 
{\em accelerator modes} (AM) resulting in super-diffusive
anomalous transport. In this paper, we describe quantum signatures of 
various classical transitions in transport characteristics emphasizing 
the cross-over effects from large to small values of effective
Planck constant $\hbar$. As $\hbar \to 0$, the quantum system exhibiting localization
in one of the phase-space directions is found to {\em feel} the effects of all classical 
transitions. However, in the absence of localized transport, the quantum system 
exhibits many surprising features and appears to be insensitive to the classical 
dynamics. 

The problem of establishing classical-quantum correspondence 
in quasi-periodic extended systems has proven to be difficult due 
to numerical limitations in approaching $\hbar \rightarrow 0$. 
All previous studies (see e.g. \cite{rev}) addressing this question have been 
limited to $\hbar \approx 1$.
Here we use recently developed {\em renormalization group} (RG) 
approach\cite{PSS} to study quantum transport for extremely small 
values of $\hbar$ upto $ \approx 10^{-4}$. 

The kicked Harper model\cite{Lima,rev} has emerged as an important model in 
quantum chaos literature. The system is given by doubly periodic 
time-dependent Hamiltonian
\begin{equation}
H(t)=L\cos(p)+K \cos(q) \sum_{k=-\infty}^\infty \delta(t-k) .
\label{KHamil}
\end{equation}
Here $q,p$ is a canonically conjugate pair of variables, usually considered
on a cylinder $p\in (-\infty,\infty)$, $q\in [0,2\pi)$.

The classical dynamics of the kicked Harper
is determined by two parameters $K$ and $L$. In the {\em asymmetric} case 
($K \ne L$), the phase space, for small values of parameters, is stratified with 
KAM tori which inhibit the transport in global scale. 
For $L > K$, or $K > L$, these tori barriers limit 
the transport along $p$, or $x$, directions, respectively.
As we show below, the 2-d parameter space
exhibits an intricately mixed non-diffusive (KAM regime) and 
diffusive regions corresponding to global stochasticity.
In contrast, in the {\em symmetric case} $K=L$, there are no KAM 
barriers to global transport but a separatrix and it is the breakup of the 
separatrix that results in global diffusion. For large values of 
the parameters, both the symmetric and the asymmetric model exhibits mostly 
diffusive behavior with the exception of narrow windows in parameter space
where the AMs  
give rise to super-diffusive transport.\cite{GZ}

The quantized system that is periodically kicked is described by 
the quasi-energy states of the one step time evolution operator, introducing 
an additional parameter $\hbar$ into the problem.
However, one hopes to recover $\hbar$ independent behavior 
(as $\hbar \to 0$) in order to establish quantum signatures of classical behavior.
It is a well established fact that the RG 
approach provides the most effective tool in distinguishing ballistic, 
diffusive and localized transport. 
Here we use recently developed\cite{PSS} dimer decimation approach to study transport
characteristics of the quasi-energy states 
in the small $\hbar$ limit. 
The RG method is applied\cite{PSS} to the 
momentum lattice ($p_m = \hbar m$) representation of the kicked model\cite{QC,Dima} 
\begin{equation}
\sum_{r=-\infty}^{\infty} B^m_r u_{m+r}= 0,
\label{Beq}
\end{equation}
where the coefficients $B^m_r$ are  
\begin{equation}
B^m_r = J_r(\bar{K}) \sin[\bar{L}\cos(m\hbar)-\pi r/2 -\omega/2].
\label{Bs}
\end{equation}
We introduce {\em renormalized parameters} as 
$\bar{K}=K/(2\hbar)$, $\bar{L}=L/(2\hbar)$. 
Tight-binding model (TBM) (\ref{Beq}) effectively contributes only 
few terms as Bessel's function exhibit fast decay when $|r| > |\bar{K}|$.
Therefore, the TBM describes a lattice model with a finite range of 
interaction denoted as $b$ ($b\approx\bar{K}$). In the limit of small 
$\bar{K},\bar{L},\omega$, TBM reduces to the Harper equation 
with $\epsilon=\hbar\omega$\cite{Harper}. 
We will choose $\hbar$ to be an irrational number with a golden tail:
$\hbar/(2\pi) = 1/(n_{\rm h} + \sigma)$ where
$\sigma =(\sqrt{5}-1)/2)$ is fixed and $\hbar$ is varied
by varying the integer $n_{\rm h}$. 
This corresponds to studying system sizes $N_n, n=1,2\ldots$ 
determined from the Fibonacci equation $N_{n+1} = N_{n} + N_{n-1}$, with 
$N_0 = 1, N_1 = n_{\rm h}$, and corresponding to $n$-th successive rational 
approximant of the irrational number $\sigma$. 
The RG methods can be used to study system sizes upto $10^9$\cite{footnote}
which allows very large $n_{\rm h}$ and hence facilitates studying kicked model for extremely small values of $\hbar$.

The transport characteristics of the quasi-energy states are studied by
computing the transmission probability $T$ on the momentum lattice. 
This is achieved in two steps: first we decimate
the lattice and then solve the scattering problem on the renormalized
lattice\cite{PSS}. Renormalization scheme makes the 
solution of the scattering problem for large lattices of size $N$ 
very efficient, as the dimer decimation reduces the size of the 
{\em scattering region}. For a fixed $\hbar$, we compute the transmission probability 
$T(N)$ for various sizes $N$ of the momentum lattice corresponding to a rational 
approximant of $\sigma$ with denominator $N$.
The scaling exponent 
$\beta = \lim_{N\to\infty}\log T(N)/\log N$ distinguishes 
extended, localized and critical states  respectively, described
by $\beta(N)$ $\to 0$, $\to -\infty$, and 
{\em oscillatory function} $\beta(N_n)$ of $n$
\cite{PSS}. 
For the exponential localization, the quantity
$\xi = -[\lim_{N\to\infty} (1/N)\log T(N)]^{-1}$ has been found to be closely
related to the localization length of the quasienery eigenstate $\omega$.

\begin{figure}
\begin{center}
\leavevmode
\epsfxsize=3in
\epsfbox{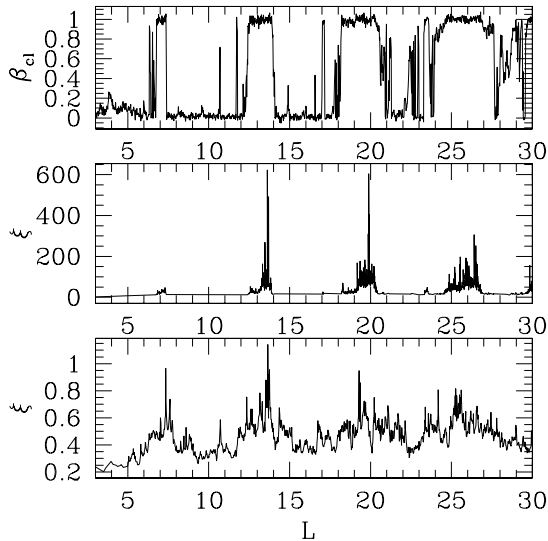}
\end{center}
\protect\caption{For fixed $\bar{K}=0.4$, 
the figure describes a series of breakup of KAM barriers 
resulting in transitions to diffusive transport $\beta_{\rm cl}=1$ (top). 
Middle and the bottom figures show the corresponding plots of quantum localization length $\xi$ for 
$n_{\rm h}=200$ and $n_{\rm h}=32$. The renormalization is carried out for system sizes increasing
until the transmission probability becomes zero. }
\label{fig1}
\end{figure}

In contrast to the kicked rotor, the kicked Harper is found to exhibit
a series of breakups and reformation of KAM barriers. 
These transitions, quantified by the exponent
$\beta_{\rm cl}=\lim_{t \to \infty}\log \langle (p(t)-p(0))^2\rangle /\log t$, are 
signaled by $\beta_{\rm cl}$ changing from $0$ to $\approx 1$.
In the quantum model, our detailed analysis for various values of $\hbar$ and $L > K$,
confirms the previously held view\cite{QC} that the quantum system remains localized in the classically diffusive regime.
However, the classical transitions corresponding to diffusive transport
manifest in huge enhancement of 
localization length.
Results for an individual pure state $\omega=0$ are shown in Fig.~1.
As demonstrated in the figure, small $\hbar$ is crucial to see signatures of $all$ classical transitions.
We would like to point out that our results are consistent with the relation
$\xi = \frac{1}{2}D/\hbar^2$ \cite{Dima}. However,
near the peaks, (narrow windows in parameter space 
corresponding to the onset to classical transitions), 
the quantum transmission probability $T(N)$ ceases to look like a simple exponential 
$\sim \exp(-N/\xi)$ 
thus making quantitative comparison difficult.

An interesting aspect of two-parameter Harper kicked model is that 
the boundary between the KAM and diffusive phases appears 
to be fractal as seen in Fig.~2. This behavior is reminiscent of the
kicked rotor problem where the kicking
potential consists of two-harmonics\cite{KM}. Although somewhat smeared, the quantum
model exhibits similar behavior: the boundary describes the transition to
the enhancement of localization length $\xi$. It is remarkable that the quantum system feels the presence
of all classical transitions and the fact that unlike kicked rotor, there is a whole hierarchy of transitions
in Harper model, makes this model an important system in quantum chaos studies.

\begin{figure}
\begin{center}
\leavevmode
\epsfxsize=3in
\epsfbox{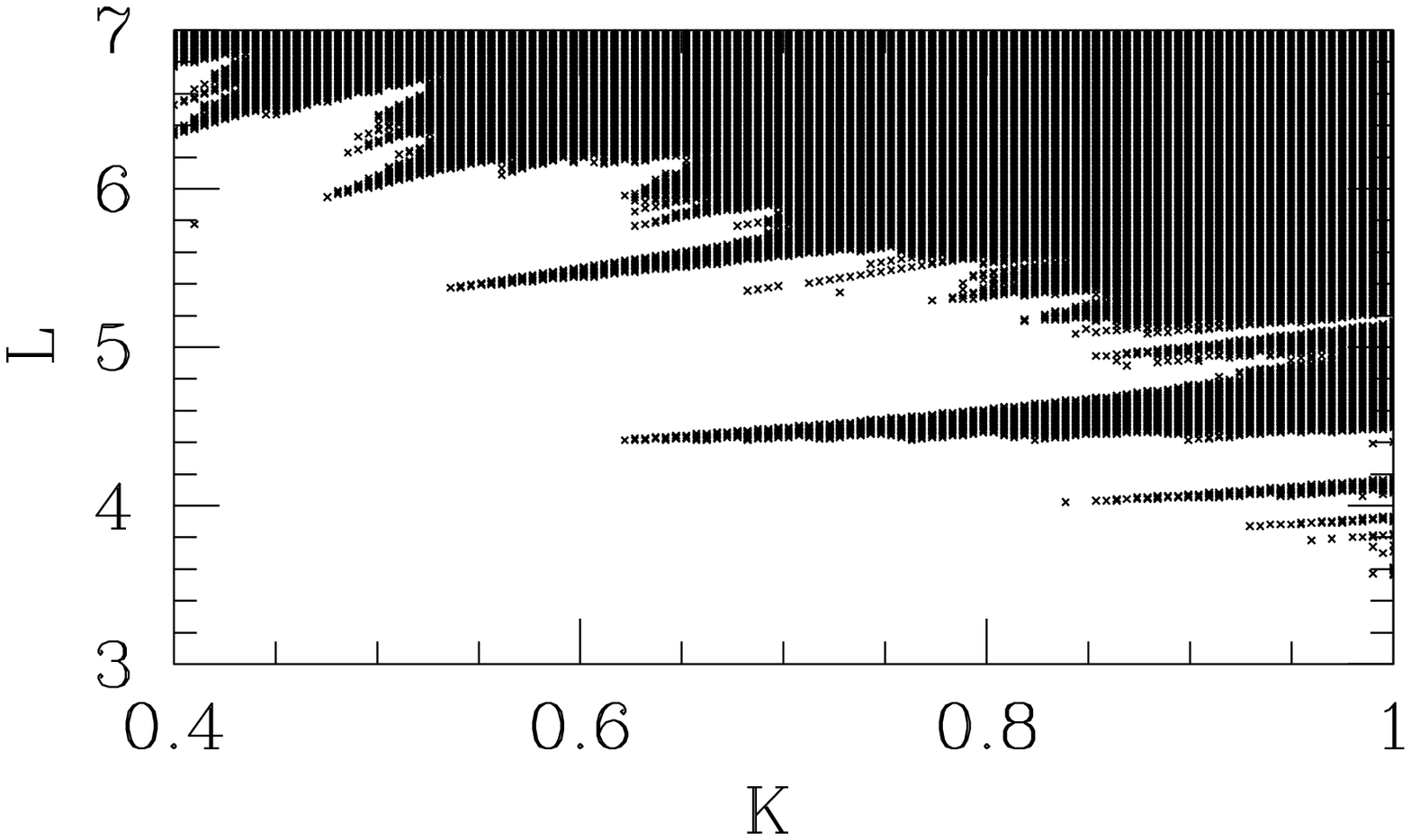}
\end{center}
\label{fig2a}
\end{figure}

\begin{figure}
\begin{center}
\leavevmode
\epsfxsize=3in
\epsfbox{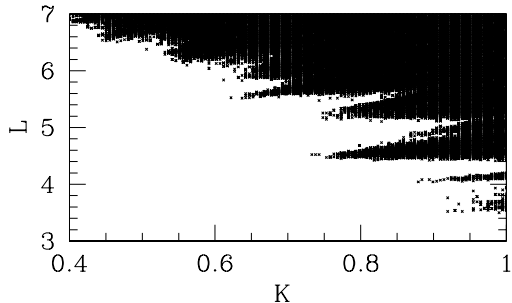}
\end{center}
\protect\caption{The diffusive regime in 2-d parameter space . 
The unshaded part(top) in the is the regime of KAM barriers for momentum transport with $\beta_{\rm cl}=0$. 
The corresponding quantum plot(bottom) with $n_{\rm h}=300$: In the unshaded part, the transport exponent 
is $\bar{\beta} \le 10$.}
\label{fig2b}
\end{figure}

Important feature of kicked systems with toroidal phase space are the AMs
which are regular (stable) space-time structures coexisting with the chaotic
sea in phase space and are accompanied by an hierarchy of island chains
inducing anomalous transport $\beta_{\rm cl} > 1$.
Fig.~3 shows one such super-diffusive parameter window whose origin is traced to a 
period-$8$ AM \cite{SS2}. 
Once again, the quantum state $\omega = 0$ although localized exhibits a very 
strong enhancement
of localization length in the classically super-diffusive regime.
It should be noted that 
in contrast to the diffusive peaks, super-diffusive spikes 
are in fact groups of many spikes 
exhibiting sensitive dependence on the parameters
and hence describe transport in fractal phase space.

\begin{figure}
\begin{center}
\leavevmode
\epsfxsize=3in
\epsfbox{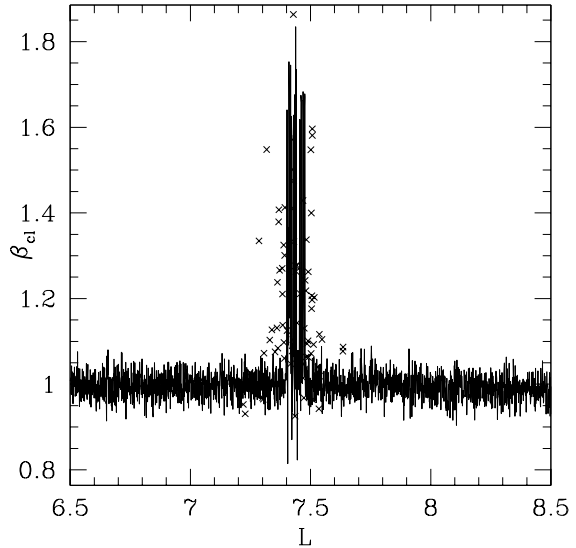}
\end{center}
%\vspace{-1.4in}
\protect\caption{For a fixed $K=0.86$ the figure shows classical anomalous transport due to AM. 
In the corresponding quantum results, the transmission probability was zero except at the crosses 
for $n_{\rm h}=300$ and the lattice size $N=43289$. More precisely, the crosses indicate points where
localization length varies between $400$ to $800$ which is about ten fold increase from the localization length
in the diffusive regime.}
\label{fig3}
\end{figure}

It is rather surprising that a pure quantum state $\omega=0$\cite{omega0} 
can exhibit such a clear signature of almost all the classical transitions.
This property may be associated with possible special structure of $\omega=0$ state
related to the fact that its quasi-energy is constant and thus insensitive to variations 
of parameters. We would like to point out that we have only investigated $L > K$ part of 
the parameter space whereas duality implies that analogous behavior will be seen for 
$K > L$ in $x$-space.
%We believe that replacing the pure state with a wave packet consisting of a superposition
%of many eigenstates may broaden the various reentrant phases as needed to sharpen the 
%classical-quantum correspondence.

In our earlier studies\cite{PSS} for $\hbar \approx O(1)$, we found 
patches of ballistic (localized) regions for $L > K$ ($K > L$)\cite{PSS}. 
Numerical studies for smaller values of $\hbar$ suggest that the overall
measure of the extended (localized) regimes for $L > K$ ($K > L$) 
approaches zero as $\hbar \rightarrow 0$.

We now discuss the symmetric Harper model with $K=L$. 
Here the quasi-energy states remain critical
and hence exhibit diffusive transport for all values of the kicking parameter $K$.
As $\hbar \rightarrow 0$, (see Fig.~4) the transmission exponent becomes $\hbar$ independent
provided we use renormalized parameter $\bar{K}$ instead of the bare $K$.
The model exhibit transmission characteristics of the Harper equation
for $\bar{K} \le \pi/2$. Precisely at $\bar{K}=\pi/2$, the transport exponent begins 
to exhibit an oscillatory behavior (with frequency proportional to $\hbar$).
As opposed to classical mechanics, where infinitesimal perturbation
leads to chaotic regions whose size increases as the perturbation increases, the 
perturbation of such quantum system causes no immediate change in the transport 
characteristics. This suggests that roots of
these transitions may be topological \cite{Leboeuf}.
It should be noted that the onset to oscillatory behavior is seen at higher odd multiples of $\pi/2$, 
however, the behavior
becomes prominent only at very small $\hbar$.
Other fascinating feature is that beyond $\bar{K} = \pi$, 
transmission probability appears to be periodic in $\bar{K}$ with period $\pi$. Finally,
the model exhibits a series of "transitions" precisely at $\bar{K} = l\pi, l=1, 2...$
characterized by a discontinuity in the transport exponent.

\begin{figure}
\begin{center}
\leavevmode
\epsfxsize=3in
\epsfbox{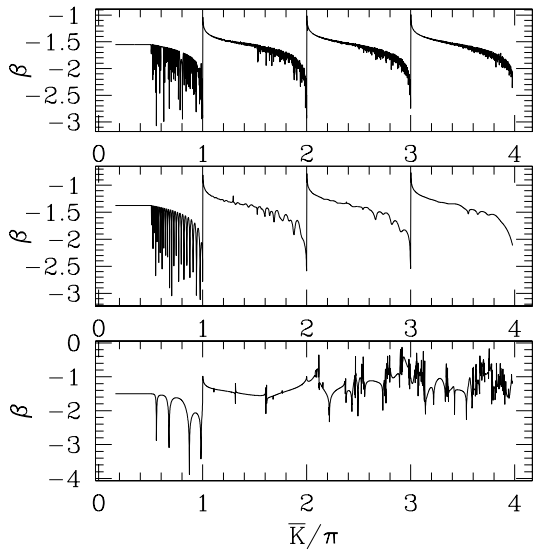}
\end{center}
\protect\caption{Variations in $\beta$ characterizing the transmission
probability along the line $K=L$ in kicked Harper model for 
$n_{\rm h}=10000$(top), $n_{\rm h}=250$(middle) and $n_{\rm h}=32$(bottom).
%The three different $\hbar$ results respectively corresponds to $N=340021,94483,220661$.
The plateau for $\bar{K} < \pi/2$ is the transport exponent for the Harper equation. 
Although the exponent $\beta$ oscillates with the $N$, all RG
iterates show qualitatively the same behavior.} 
\label{fig4}
\end{figure}

Recently, a semiclassical analysis\cite{SS} hinted a possibility of a series of 
enhancement of transport at $\bar{K} = l \pi/2$ associated with
the emergence of new periodic orbits 
of frequency $l/4$. The integer $l$ is a kind of winding number that 
unfolds $\omega$ (confined to the interval $0-2\pi$).
The fact that $\bar{K}$ and not $K$ determines the thresholds for various 
transitions poses a serious problem about the classical-quantum correspondence. 
As discussed in earlier studies\cite{SS} the symmetric model exhibits super-diffusive 
classical transport at various critical values of the kicking parameter $K$. 
The most important of those are the 
period-1 and period-2 AMs which respectively occur at even and odd multiples
of $\pi$. It is therefore tempting to associate $\bar{K}$ with the classical $K$ as was 
suggested in Ref.\cite{SS}. 
However, this singular scaling which associates the 
discontinuities in Fig.~4 with the classical
AMs is inconsistent with
the asymmetric case where no scaling of the
parameters is needed to establish quantum-classical correspondence.
It is possible that the transitions
seen in the symmetric case are purely quantum-mechanical in nature
and may have their origin in resonances of the (driven) RG flow 
and/or topological changes.
We should mention that a possibility of some sudden changes at multiples of $\pi/2$ also
emerged in our analysis of the scattering problem. It turns out that
the number of independent propagating solutions of the
scattering problem\cite{PSS}, i.e. dimension of the S-matrix,
%is just $N_{\rm o} = [2\bar{K}/\pi]$, where $[x]$ is the closest integer to $x$. $N_{\rm o}$
increases by $1$ at $\bar{K}_l = l \pi/2,l=1,2\ldots$, matching with the discontinuities of
the transport exponent $\beta(\bar{K})$.
Finally, the most challenging result which defies our intuition, is the
(asymptotic) periodicity of the transmission probability 
$T|_{\bar{K}+\pi} = T|_{\bar{K}}$, for $\bar{K}\gg \pi$
%If this periodicity continues to persist for large bare $K$ values, i.e.
%$\bar{K}\gg \hbar^{-1}$ (a regime inaccessible to
%RG analysis for $\hbar \to 0$\cite{footnote}) 
which rules out any possibility
of quantum manifestations of classical super-diffusive transitions.

Symmetric Kicked Harper model is an interesting example of a non-integrable system where
the classical as well as the quantum transport is diffusive. 
It is in sharp contrast to the asymmetric case where the classically diffusive behavior
corresponds to localized quantum transport. In view of this, it is rather surprising that 
in the asymmetric case the quantum system appears to respond to {\em all} the changes in the 
classical behavior, while in the symmetric case it remains insensitive to the 
variation in classical transport and instead repeats its behavior 
at every multiple of $\pi$. This adds a new puzzle to the field of
quantum chaos.

Inability of the quantum system to delocalize in classically diffusive regime and mimic the classical behavior for 
arbitrary small value of $\hbar$ as confirmed by RG analysis,
remains an open frontier. Earlier studies have suggested phase 
randomization\cite{QC} due to classical chaos as a 
mechanism for quantum dynamical localization.  
The fact that the kicked Harper model can 
exhibit localized, ballistic and diffusive 
transport irrespective of the fact that
the corresponding classical system is chaotic challenges the phase 
randomization as the underlying
mechanism for localization.
Here we would like to propose an alternative mechanism of dynamical 
localization: 
we speculate that the dynamical localization may be due to the 
cantori barriers. 
These are invariant quasiperiodic trajectories with infinite number of steps 
and provide an effective non-analytic quasiperiodic potential. 
The possibility of localization in 
quasiperiodic potential with {\it infinite steps}  
has been discussed recently\cite{KSlong}.
We would like to emphasize that the scenario for localization
due to cantori suggests that these barriers continue to inhibit transport
even as $\hbar \to 0$ irrespectively of the flux through the holes in cantori.
This scenario not only explains dynamical localization in kicked rotor and 
Harper model (for $L > K$),
but also accounts for ballistic transport in kicked Harper for $K > L$. Furthermore,
it is consistent with the diffusive quantum transport for $K=L$ case
, since the symmetric Kicked Harper model does not possess
global cantori barriers (at least not of the type of broken KAM barriers).
We hope that further studies will put our speculative views on solid footings.

The research of IIS is supported by National Science
Foundation Grant No. DMR~0072813. TP acknowledges Ministry of Education, Science and Sport
of Slovenia for a financial support.


\begin{references}
\bibitem{omega0} To our knowledge, the interesting observation, namely that $\omega=0$ is
always in the spectrum of Kicked Harper model, has not been proved.
\end{references}
\end{document}